\documentclass[MSNbibl,nameyear,dvips]{arxstspdf}
\usepackage{flushend}
\usepackage{stfloats}
\usepackage{graphicx}

\volume{28}
\issue{2}
\pubyear{2013}
\firstpage{209}
\lastpage{222}
\doi{10.1214/12-STS407} %kopijuoti is PTS
\makeatletter

\newtheorem{lemma}{Lemma}
\newtheorem{theorem}{Theorem}
\newtheorem{corollary}{Corollary}

\newcommand{\citecs}[1]{\citeauthor{#1}, \citeyear{#1}}
\renewcommand{\citep}[1]{(\citeauthor{#1}, \citeyear{#1})}

\newcommand{\iid}{\stackrel{\mathrm{i.i.d.}}{\sim}}

\newcommand{\bP}{{\mathbf P}}
\newcommand{\bs}{{\mathbf s}}

\newcommand{\Pb}{\bP}
\newcommand{\Pbell}{\Pb^{(\ell)}}
\newcommand{\Pell}{P^{(\ell)}}
\newcommand{\tPb}{\tilde{\bP}}
\newcommand{\tPbell}{\tilde{\Pb}^{(\ell)}}
\newcommand{\pibell}{{\pi}^{(\ell)}}
\newcommand{\piell}{\pi^{(\ell)}}

\newcommand{\phat}{\widehat{p}}

\newcommand{\cX}{\mathcal{X}}

\newcommand{\bbN}{\mathbb{N}}
\newcommand{\bbP}{\mathbb{P}}

\newcommand{\bfn}{\mathbf{n}}
\newcommand{\bfone}{\mathbf{1}}

\newcommand{\Xt}{\tilde{X}}
\newcommand{\Fbar}{\overline{F}}

\newcommand{\dd}{\,\mathrm{d}}

\newcommand{\Ga}{\operatorname{Ga}}

\renewcommand{\th}{\theta}
\newcommand{\SSM}{\operatorname{SSM}}
\newcommand{\DP}{\operatorname{DP}}

\makeatother

\begin{document}
\begin{frontmatter}

\title{Defining Predictive Probability Functions for Species
Sampling Models}%\thanksref{T1}
% kai straipsnis turi susijusiu diskusiju ir rejoinder'iu
%rejoinder at \relateddoi{r}{10.1214/00-STSXXXX}.}
\runtitle{Predictive Probability Functions}

\begin{aug}
\author[a]{\fnms{Jaeyong} \snm{Lee}\ead[label=e1]{leejyc@gmail.com}},
\author[b]{\fnms{Fernando A.} \snm{Quintana}\corref{}\ead[label=e2]{quintana@mat.puc.cl}},
\author[c]{\fnms{Peter} \snm{M\"uller}\ead[label=e3]{pmueller@math.utexas.edu}}
\and
\author[d]{\fnms{Lorenzo} \snm{Trippa}\ead[label=e4]{ltrippa@jimmy.harvard.edu}}
\runauthor{Lee, Quintana, M\"uller and Trippa}

\affiliation{Seoul National University, Pontificia Universidad Cat\'
olica de Chile,
University of Texas at Austin and Harvard University}

\address[a]{Jaeyong Lee is Professor, Department of Statistics,
Seoul National University, Seoul, 151-747, South Korea
\printead{e1}.}
\address[b]{Fernando A. Quintana is Professor,
Departamento de Estad\'{\i}stica,
Pontificia Universidad Cat\'olica de Chile,
Macul, Santiago 22, Chile \printead{e2}.}
\address[c]{Peter M\"uller is Professor,
Department of Mathematics,
University of Texas at Austin,
Austin, Texas 78712-1202, USA \printead{e3}.}
\address[d]{Lorenzo Trippa is Assistant Professor,
Department of Biostatistics, Harvard University,
and
Department of Biostatistics and Computational  Biology,
Dana-Farber Cancer Institute, Boston, Massachusetts
02115, USA
\printead{e4}.}

\end{aug}

% ABSTRACT
%
\begin{abstract}
We review the class of species sampling models (SSM). In particular, we
investigate the relation between the exchangeable partition probability
function (EPPF) and the predictive probability function (PPF). It is
straightforward to define a PPF from an EPPF, but the converse is not
necessarily true. In this paper we introduce the notion of putative
PPFs and show novel conditions for a putative PPF to define an EPPF. We
show that all possible PPFs in a certain class have to define
(unnormalized) probabilities for cluster membership that are linear in
cluster size. We give a new necessary and sufficient condition for
arbitrary putative PPFs to define an EPPF. Finally, we show posterior
inference for a large class of SSMs with a PPF that is not linear in
cluster size and discuss a numerical method to derive its PPF.
\end{abstract}

% KEYWORDS
%

\end{frontmatter}

%s1 #&#
\section{Introduction}
\label{secintro}

The status of the Dirichlet process (\citeauthor{Ferguson73},\break \citeyear{Ferguson73})
(DP) among
nonparametric priors is comparable to that of the normal distribution
among finite-dimensional distributions. This is in part due to the
marginalization property: a random sequence sampled from a random
probability measure with a Dirichlet process prior forms marginally a
Polya urn sequence \citep{Blackwell1973}. Markov chain Monte Carlo
simulation based on the marginalization property has been the central
computational tool for the DP and facilitated a wide variety of
applications. See \citet{MacEachern94}, \citet
{EscobarWest95} and
\citet{MacEachernMuller98}, to name just a few. In Pitman
(\citeyear{Pitman95,Pitman96}), the species sampling model
(SSM) is proposed as a generalization of the DP. SSMs can be used as
flexible alternatives to the popular DP model in nonparametric Bayesian
inference. The SSM is defined as the directing random probability
measure of an exchangeable species sampling sequence which is defined
as a generalization of the Polya urn sequence. The SSM has a
marginalization property similar to the DP. It therefore enjoys the
same computational advantage as the DP while it defines a much wider
class of random probability measures. For its theoretical properties
and applications, we refer to \citet{IshwaranJames03},
\citet{LijoiMena05}, \citet{LijoiPrunster05}, \citet{James08},
\citet{NavarreteQuintana08}, \citet{JamesLijoi09} and
\citet{JangLee10}.

Suppose $(X_1,X_2,\ldots)$ is a sequence of random variables. In a
traditional application the sequence arises as a random sample from a
large population of units, and $X_i$ records the species of the $i$th
individual in the sample. This explains the name SSM. Let $\Xt_j$ be
the $j$th distinct species to appear. Let $n_{jn}$ be the number of
times the $j$th species $\Xt_j$ appears in $(X_1,\ldots,X_n)$,
$j=1,2,\ldots\,$, and
\[
\bfn_n = %(n_{1n}, n_{2n},\ldots) \mbox{ or }
(n_{jn}, j=1,\ldots,k_n),
\]
where $k_n = k_n(\bfn_n) = \max\{j\dvtx n_{jn} > 0 \}$ is the number of
different species to appear in $(X_1,\ldots,X_n)$. The sets $\{i\le
n\dvtx X_i=\Xt_j\}$ define clusters that partition the index set
$\{1,\ldots,n\}$. When $n$ is understood from the context we just write
$n_j$, $\bfn$ and $k$ or $k(\bfn)$.

We now give three alternative characterizations of species sampling
sequences: (i) by the predictive probability function, (ii) by the
driving measure of the exchangeable sequence, and (iii) by the
underlying exchangeable partition probability function.

\subsection*{PPF}
Let $\nu$ be a diffuse (or nonatomic) probability measure on a complete
separable metric space $\cX$ equipp\-ed with Borel $\sigma$-field. An
exchangeable sequence $(X_1,\break X_2,\ldots)$ is called a species sampling
sequence (SSS) if $X_1 \sim\nu$ and
%
%e1 #&#
\begin{eqnarray}\label{eqpred}
&&
X_{n+1}\mid X_1,\ldots,X_n \nonumber\\[-8pt]\\[-8pt]
&&\quad\sim\sum
_{j=1}^{k_n} p_j(\bfn_n)
\delta_{\Xt_j} + p_{k_n+1}(\bfn_n) \nu,\nonumber
\end{eqnarray}
where $\delta_x$ is the degenerate probability measure at~$x$. Examples
of SSS include the P\'olya urn sequence $(X_1, X_2, \ldots)$ whose
distribution is the same as the marginal distribution of independent
observations from a Dirichlet random distribution $F$, that is, $ X_1,\break X_2,
\ldots\mid F \iid F$ with $ F \sim \DP(\alpha
\nu)$, where $\alpha> 0$. The conditional distribution of the P\'olya
urn sequence is
\[
X_{n+1}\mid X_1,\ldots,X_n \sim\sum
_{j=1}^{k_n} \frac{n_j}{n+\alpha} \delta_{\Xt_j} +
\frac{\alpha}{n+\alpha} \nu.
\]
This marginalization property has been a central tool for posterior
simulation in DP mixture models, which benefit from the fact that one
can integrate out $F$ using the marginalization property. The posterior
distribution becomes then free of the infinite-dimensional object $F$.
Thus, Markov chain Monte Carlo algorithms for DP mixtures do not pose
bigger difficulties than the usual parametric Bayesian models
(\citecs{MacEachern94}; \citeauthor{MacEachernMuller98},\break \citeyear{MacEachernMuller98}).
Similarly, alternative\vadjust{\goodbreak} discrete random distributions have been considered in the literature
and proved computationally attractive due to analogous marginalization
properties; see, for example, Lijoi, Mena and Pr{\"u}nster
(\citeyear{LijoiMena05,Lijoi2007}).

The sequence of functions $(p_1,p_2,\ldots)$ in (\ref{eqpred}) is
call\-ed a sequence of predictive probability\vspace*{1pt}
functions\break
(PPF). These are defined on
$\bbN^*=\bigcup_{k=1}^\infty\bbN^k$, where $\bbN$ is the set of natural
numbers, and satisfy the conditions
%
%e2 #&#
\begin{eqnarray}
\label{pppf}
&&p_j(\bfn) \geq0 \quad\mbox{and}\quad \sum
_{j=1}^{k_n+1} p_j(\bfn) = 1\nonumber\\[-8pt]\\[-8pt]
&&\eqntext{\mbox{for all } \bfn\in\bbN^*.}
\end{eqnarray}
Motivated by these properties of PPFs, we define a sequence of
\textit{putative PPFs} as a sequence of functions $(p_j,j=1,2,\ldots)$ defined
on $\bbN^*$ which satisfies~(\ref{pppf}). Note that not all putative
PPFs are PPFs, because (\ref{pppf}) does not guarantee exchangeability
of $(X_1,X_2,\ldots)$ in~(\ref{eqpred}). Note that the weights
$p_j(\cdot)$ depend on the data only indirectly through the cluster
sizes $\bfn_n$. The widely used DP is a special case of a species
sampling model, with $p_j(\bfn_n) \propto n_j$ and $p_{k+1}(\bfn_n)
\propto\alpha$ for a DP with total mass parameter $\alpha$. The use of
$p_j$ in (\ref{eqpred}) implies
\begin{eqnarray}
p_j(\bfn) &=& \bbP(X_{n+1} = \Xt_j \mid
X_1,\ldots,X_n),\nonumber\\
&&\eqntext{j=1,\ldots,k_n,}
\\
p_{k_n+1}(\bfn) & = & \bbP\bigl(X_{n+1} \notin\{ \tilde
X_1,\ldots, \tilde X_{k_n}\} \mid X_1,
\ldots,X_n\bigr).\nonumber
\end{eqnarray}
In words, $p_j$ is the probability of the next observation being the
$j$th species (falling into the $j$th cluster) and $p_{k_n+1}$ is the
probability of a new species (starting a new cluster).

An important point in the above definition is that a sequence $X_i$
can be a SSS only if it is exchangeable.

\subsection*{SSM}
Alternatively, a SSS can be characterized by the following defining
property. An exchangeable se-\break quence of random variables
$(X_1,X_2,\ldots)$ is a species sampling sequence if and only if
$X_1,X_2,\ldots\mid G$ is a random sample from $G$ where
%
%e3 #&#
\begin{equation}
\label{eqssp} G = \sum_{h=1}^\infty
P_h \delta_{m_h} + R\nu
\end{equation}
for some sequence of positive random variables $(P_h)$ and $R$ such
that $1-R = \sum_{h=1}^\infty P_h \leq1$ with probability 1, $(m_h)$
is a sequence of independent variables with distribution\vadjust{\goodbreak} $\nu$,
and $(P_i)$ and $(m_h)$ are independent. See \citet{Pitman96}. The
result is an extension of de Finetti's theorem and characterizes
the directing random probability measure of the species sample
sequence. We call the directing random probability measure $G$ in
equation (\ref{eqssp}) the \textit{SSM} of the\break SSS $(X_i)$.

\subsection*{EPPF}
A third alternative definition of a SSS and corresponding SSM is in
terms of the implied probability model on a sequence of random
partitions.

Suppose a SSS $(X_1, X_2, \ldots)$ is given. Since the de Finetti
measure (\ref{eqssp}) is partly discrete, there are ties among
$X_i$'s. The ties among $(X_1,X_2, \ldots, X_n)$ for a given $n$ induce
an equivalence relation in the set $[n]=\{1,2,\ldots,n\}$, that is, $i
\sim j$ if and only if $X_i = X_j$. This equivalence relation on
$[n]$, in turn, induces the partition $\Pi_n$ of $[n]$. Due to the
exchangeability of $(X_1, X_2, \ldots)$, it can be easily seen that the
random partition $\Pi_n$ is an exchangeable random partition on $[n]$,
that is, for any partition $\{A_1, A_2, \ldots, A_k \}$ of $[n]$, the
probability $P(\Pi_n = \{A_1, A_2, \ldots, A_k \})$ is invariant under
any permutation on $[n]$ and can be expressed as a function of $\bfn=
(n_1, n_2, \ldots, n_k)$, where $n_i$ is the cardinality of $A_i$ for
$i=1,2,\ldots,k$. Extending the above argument to the entire SSS, we
can get an exchangeable random partition on the natural numbers $\bbN$
from the SSS. Kingman (\citeyear{Kingman78,Kingman82})
showed a remarkable result, called Kingman's representation theorem,
that in fact every exchangeable random partition can be obtained by a
SSS.

For any partition $\{A_1, A_2, \ldots, A_k \}$ of $[n]$, we can
represent $P(\Pi_n = \{A_1, A_2, \ldots, A_k \}) = p(\bfn)$ for a
symmetric function $p\dvtx \bbN^* \rightarrow[0,1]$ satisfying
%
%e4 #&#
\begin{eqnarray}
\label{eqeppf} p(1) &=& 1,
\nonumber\\[-8pt]\\[-8pt]
p(\bfn) &=& \sum_{j=1}^{k(\bfn)+1}p\bigl(
\bfn^{j+}\bigr)\qquad\mbox{for all } \bfn\in\bbN^*,
\nonumber
\end{eqnarray}
where $\bfn^{j+}$ is the same as $\bfn$ except that the $j$th element
is increased by $1$. This function is called an exchangeable partition
probability function (EPPF) and characterizes the distribution of an
exchangeable random partition on $\bbN$.

We are now ready to pose the problem for the present paper. It is
straightforward to verify that any EPPF defines a PPF by
%
%e5 #&#
\begin{equation}
\label{eqrelation} p_j(\bfn) = \frac{p(\bfn^{j+})}{p(\bfn)},\qquad
j=1,2,\ldots,k+1.
\end{equation}
The converse is not true. Not every putative $p_j(\bfn)$ defines an
EPPF and thus a SSM and a SSS.\vadjust{\goodbreak} For example, it is easy to show that
$p_j(\bfn) \propto n_j^2+1$, $j=1,\ldots,k(\bfn)$, does not. In Bayesian
data analysis it is often convenient, or at least instructive, to
elicit features of the PPF rather than the joint EPPF. Since the PPF is
crucial for posterior computation, applied Bayesians tend to focus on
it to specify the species sampling prior for a specific problem. For
example, the PPF defined by a DP prior implies that the probability of
joining an existing cluster is proportional to the cluster size. This
is not always desirable. Can the user define an alternative PPF that
allocates new observations to clusters with probabilities proportional
to alternative functions $f(n_j)$ and still define a SSS? In general,
the simple answer is no. We already mentioned that a PPF implies a SSS
if and only if it arises as in (\ref{eqrelation}) from an EPPF. But
this result is only a characterization. It is of little use for data
analysis and modeling since it is difficult to verify whether or not a
given PPF arises from an EPPF. In this paper we develop some conditions
to address this gap. We consider methods to define PPFs in two
different directions. First we give an easily verifiable necessary
condition for a putative PPF to arise from an EPPF (Lemma~\ref{lemm1}) and a
necessary and sufficient condition for a putative PPF to arise from an
EPPF. A consequence of this result is an elementary proof of the
characterization of all possible PPFs with form $p_j(\bfn) \propto
f(n_j)$. This result has been proved earlier by \citet{GnedinPitman06}.
Although the result in Section~\ref{secppfeppf} gives necessary and sufficient
conditions for a putative PPF to be a PPF, the characterization is not
constructive. It does not give any guidance in how to create a new PPF
for a specific application. In Section~\ref{seclorenzo} we propose an alternative
approach to define a SSM based on directly defining a joint probability
model for the $P_h$ in (\ref{eqssp}). We develop a numerical algorithm
to derive the corresponding PPF. This facilitates the use of such
models for nonparametric Bayesian data analysis. This approach can
naturally create PPFs with very different features than the well-known
PPF under the DP.

The literature reports some PPFs with closed-form analytic expressions
other than the PPF under the DP prior. There are a few directions which
have been explored for constructing extensions of the DP prior and
deriving PPFs. The normalization of complete random measures (CRM) has
been proposed in \citet{Kingman75}. A CRM such as the generalized
gamma process \citep{Brix1999}, after normalization, defines a discrete
random distribution and, under mild assumptions, a SSM. Developments\vadjust{\goodbreak}
and theoretical results on this approach have been discussed in a
series of papers; see, for example, \citet{PermanPitman92},
\citet{Pitman03} and \citet{RegazziniLijoi03}. Normalized
CRM models
have also been studied and applied in \citet{LijoiMena05},
\citet{NietoPrunster04} and more recently in \citet
{JamesLijoi09}. A
second related line of research considered the so-called Gibbs models.
In these models the analytic expressions of the PPFs share similarities
with the DP model. An important example is the Pitman--Yor process.
Contributions include \citet{Gnedin2006}, \citet{Lijoi2007},
Lijoi, Pr{\"u}nster and Walker (\citeyear{Lijoi08,Lijoi08a})
and \citet{Gnedin2010}. \citet{Lijoi2010} provide a recent
overview on
major results from the literature on normalized CRM and Gibbs-type
partitions.\looseness=1

%s2 #&#
\section{When Does a PPF Imply an EPPF?}
\label{secppfeppf}

Suppose we are given a putative PPF $(p_j)$. Using equation
(\ref{eqrelation}), one can attempt to define a function
$p\dvtx\bbN^*\rightarrow[0,1]$ inductively by the following mapping:
%
%e6 #&#
\begin{eqnarray}
\label{eqeppfdef}
p(1) &=& 1,
\nonumber\\
p\bigl(\bfn^{j+}\bigr) &=& p_j(\bfn) p(\bfn)\qquad\qquad\qquad\\
&&\eqntext{\mbox{for
all } \bfn\in\bbN\mbox{ and } j =1,2,\ldots,k(\bfn)+1.}
\end{eqnarray}
In general, equation (\ref{eqeppfdef}) does not lead to a unique
definition of $p(\bfn)$ for each $\bfn\in\bbN^*$. For example, let
$\bfn= (2,1)$. Then, $p(2,1)$ could be computed in two different ways
as $p_2(1)p_1(1,1)$ and $p_1(1)p_2(2)$ which correspond to partitions
$\{\{1,3\},\{2\}\}$ and $\{\{1,2\},\break\{3\}\}$, respectively. If
$p_2(1)p_1(1,1) \neq p_1(1)p_2(2)$, equation (\ref{eqeppfdef}) does
not define a function $p\dvtx\bbN^* \rightarrow[0,1]$. The following
lemma shows a condition for a PPF for which equation (\ref{eqeppfdef})
leads to a valid unique definition of $p\dvtx\bbN^* \rightarrow[0,1]$.

Suppose $\Pi= \{A_1,A_2,\ldots,A_k\}$ is a partition of $[n]$ with
clusters indexed in the order of appearance. For $1 \leq m \leq n$, let
$\Pi_m$ be the restriction of $\Pi$ on $[m]$. Let $\bfn(\Pi) =
(n_1,\ldots,n_k)$, where $n_i$ is the cardinality of $A_i$, and let
$\Pi(i)$
be the class index of element $i$ in partition $\Pi$ and $\Pi([n]) =
(\Pi(1),\ldots,\Pi(n))$.
%
%le1 #&#
\begin{lemma}\label{lemm1}
If and only if a putative PPF $(p_j)$ satisfies
%
%e7 #&#
\begin{eqnarray}
\label{eqsuffdef} p_i(\bfn) p_j\bigl(\bfn^{i+}
\bigr) = p_j(\bfn) p_i\bigl(\bfn^{j+}\bigr)\nonumber\\[-8pt]\\[-8pt]
&&\eqntext{\mbox{for all } \bfn\in\bbN^*, i, j =1,2,\ldots,k(\bfn)+1,}
\end{eqnarray}
then $p$ defined by (\ref{eqeppfdef}) is a function from $\bbN^*$ to
$[0,1]$, that is, $p$ in (\ref{eqeppfdef}) is uniquely defined.\vadjust{\goodbreak}
\end{lemma}
\begin{pf}
Let $\bfn= (n_1,\ldots,n_k)$ with $\sum_{i=1}^k
n_i = n$ and $\Pi$ and $\Omega$ be two partitions of $[n]$ with
$\bfn(\Pi) = \bfn(\Omega) = \bfn$. Let $p^\Pi(\bfn) = \prod_{i=1}^{n-1}
p_{\Pi(i+1)}(\bfn(\Pi_i))$ and\break $p^\Omega(\bfn) = \prod_{i=1}^{n-1}
p_{\Omega(i+1)}(\bfn(\Omega_i))$. We need to show that
$p^\Pi(\bfn)=p^\Omega(\bfn)$. Without loss of generality, we can assume
$\Pi([n]) = (1,\ldots,1,2,\ldots,2,\ldots,k,\ldots,k)$, where $i$ is
repeated $n_i$ times for $i=1,\ldots,k$. Note that $\Omega([n])$ is
just a certain permutation of $\Pi([n])$ and by a finite times of
swapping two consecutive elements in $\Omega([n])$, one can change
$\Omega([n])$ to $\Pi([n])$. Thus, it suffices to show when
$\Omega([n])$ is different from $\Pi([n])$ in only two consecutive
positions. But, this is guaranteed by condition (\ref{eqsuffdef}).

The opposite is easy to show. Assume $p_j$ defines a unique $p(\bfn)$.
Consider (\ref{eqsuffdef}) and multiply on both sides with $p(\bfn)$.
By assumption, we get on either side $p(\bfn^{i+j+})$. This completes
the proof.
\end{pf}

Note that the conclusion of Lemma~\ref{lemm1} is not (yet) that $p$ is an EPPF.
The missing property is exchangeability, that is, invariance of $p$ with
respect to permutations of the group indices $j=1,\ldots,k(\bfn)$. When
the function $p$, recursively defined by expression
(\ref{eqeppfdef}), satisfies the balance imposed by equation
(\ref{eqsuffdef}) it is called the \textit{partially exchangeable probability
function} (Pitman, \citeyear{Pitman95,Pitman06}) and
the resulting random partition of $\mathbb{N}$ is termed partially
exchangeable. In \citet{Pitman95}, it is proved that a $p\dvtx
\mathbb{N}^*\rightarrow[0,1]$ is a \textit{partially exchangeable probability
function} if and only if it exists as a sequence of nonnegative random
variables $P_i$, $i=1,\ldots\,$, with $\sum_i P_i\le1$ such that
%
%e8 #&#
\begin{equation}
\label{eqpartialcar} p(n_1,\ldots,n_k)= E \Biggl[ \prod
_{i=1}^k P_i^{n_i-1}
\prod_{i=1}^{k-1} \Biggl(1-\sum
_{j=1}^{i} P_i \Biggr) \Biggr],\hspace*{-24pt}
\end{equation}
where the expectation is with respect to the distribution of the
sequence $(P_i)$. We refer to \citet{Pitman95} for an extensive
study of
partially exchangeable random partitions.

It is easily checked whether or not a given PPF satisfies the condition
of Lemma~\ref{lemm1}. Corollary~\ref{coro1} describes all possible PPFs that have the
probability of cluster memberships depend on a function of the cluster
size only. This result is part of a theorem in \citet{GnedinPitman06},
but we give here a more straightforward proof.
%
%co1 #&#
\begin{corollary}\label{coro1}
Suppose a putative PPF $(p_j)$ satisfies (\ref{eqsuffdef}) and
%
%e9 #&#
\begin{equation}
\label{ppf1} p_j(n_1,\ldots,n_k) \propto
\cases{ %
f(n_j), & $j=1,\ldots,k$,\cr
\theta, & $j= k+1$,}
\end{equation}
where $f(k)$ is a function from $\bbN$ to $(0,\infty)$ and $\theta>
0$. Then, $f(k) = a k$ for all $k \in\bbN$ for some $a > 0$.\vadjust{\goodbreak}
\end{corollary}
\begin{pf}
Note that for any $\bfn= (n_1,\ldots,n_k)$ and
$i=1,\ldots,k+1$,
\[
p_i(n_1,\ldots,n_k) = \cases{
\dfrac{f(n_i)}{\sum_{u=1}^k f(n_u) + \theta}, & $i=1,\ldots,k$,\vspace*{2pt}\cr
\dfrac{\theta}{\sum_{u=1}^k f(n_u) + \theta}, & $i = k+1$.}
\]

Equation (\ref{eqsuffdef}) with $1 \leq i \neq j \leq k$ implies
\begin{eqnarray*}
&&\frac{f(n_i)}{\sum_{u=1}^k f(n_u) + \theta} \frac{f(n_j)}{\sum_{u
\neq i}^k f(n_u) + f(n_i+1)
+\theta} \\
&&\quad=\frac{f(n_j)}{\sum_{u=1}^k f(n_u) + \theta} \frac{f(n_i)}{\sum_{u
\neq j}^k f(n_u) +
f(n_j+1) +\theta},
\end{eqnarray*}
which in turn implies
\[
f(n_i) + f(n_j+1) = f(n_j) +
f(n_i + 1)
\]
or
\[
f(n_j+1)-f(n_j) = f(n_i+1) -
f(n_i).
\]
Since this holds for all $n_i$ and $n_j$, we have for all $k \in\bbN$
%
%e10 #&#
\begin{equation}
\label{eq1} f(m) = a m + b
\end{equation}
for some $a, b \in\mathbb{R}$.

Now consider $i=k+1$ and $1\leq j \leq k$. Then,
\begin{eqnarray*}
&&
\frac{\theta}{\sum_{u=1}^k f(n_u) + \theta} \frac{f(n_j)}{\sum_{u
=1}^k f(n_u) + f(1)
+\theta} \\
&&\quad=\frac{f(n_j)}{\sum_{u=1}^k f(n_u) + \theta} \frac{\theta}{\sum
_{u \neq j}^k f(n_u) +
f(n_j+1) +\theta},
\end{eqnarray*}
which implies $f(n_j) + f(1) = f(n_j +1)$ for all $n_j$. This together
with (\ref{eq1}) implies $b= 0$. Thus, we have $f(k) = ak$ for some $a
> 0$.
\end{pf}

For any $a > 0$, the putative PPF
\[
p_i(n_1,\ldots,n_k) \propto\cases{
a n_i, & $i=1,\ldots,k$,\cr
{\theta}, & $i = k+1$,}
\]
defines a function $p\dvtx\bbN\rightarrow[0,1]$,
\[
p(n_1,\ldots,n_k) = \frac{\theta^{k-1} a^{n-k}}{[\theta+1]_{n-1;a}}
\prod
_{i=1}^k (n_i-1)!,
\]
where $[\theta]_{k;a} = \theta(\theta+a)
\cdots(\theta+ (k-1)a)$. Since this function is symmetric in its
arguments, it is an EPPF. This is the EPPF for a DP with total mass
$\theta/a$. Thus, Corollary~\ref{coro1} implies that the EPPF under the DP is the
only EPPF that satisfies (\ref{ppf1}). The corollary shows that it is
not an entirely trivial matter to come up with a putative PPF that
leads to a valid EPPF. A version of Corollary~\ref{coro1} is also well known as
Johnson's Sufficientness postulate \citep{good65}. See also the
discussion in \citet{zabell82}.

We now give a necessary and sufficient condition for the function $p$
defined by (\ref{eqrelation}) to be an EPPF, without any constraint on
the form of $p_j$ (as were present in the earlier results). Suppose
$\sigma$ is a permutation of $[k]$ and $\bfn=(n_1,\ldots,n_k) \in
\bbN^*$. Define $\sigma(\bfn) = \sigma(n_1,\ldots,n_k) =
(n_{\sigma(1)},n_{\sigma(2)},\ldots,n_{\sigma(k)})$. In\break words,
$\sigma$
is a permutation of group labels and $\sigma(\bfn)$ is the
corresponding permutation of the group sizes~$\bfn$.

%th1 #&#
\begin{theorem}\label{theo1}
Suppose a putative PPF $(p_j)$ satisfies (\ref{eqsuffdef}) as well
as the following condition: for all $\bfn=(n_1,\ldots,n_k) \in\bbN^*$,
and permutations $\sigma$ on $[k]$ and $i=1,\ldots,k$,
%
%e11 #&#
\begin{eqnarray}
\label{eqsuffeppf}
p_i(n_1,\ldots,n_k)
=
p_{\sigma^{-1}(i)}(n_{\sigma(1)},n_{\sigma(2)},\ldots,n_{\sigma(k)}).\hspace*{-35pt}
\end{eqnarray}
Then, $p$ defined by (\ref{eqeppfdef}) is an EPPF. The condition is
also necessary; if $p$ is an EPPF, then (\ref{eqsuffeppf}) holds.
\end{theorem}
\begin{pf}
Fix $\bfn=(n_1,\ldots,n_k) \in\bbN^*$ and a
permutation on $[k]$, $\sigma$. We wish to show that for the function
$p$ defined by (\ref{eqeppfdef})
%
%e12 #&#
\begin{equation}\label{eq9a}\quad
p(n_1,\ldots,n_k) = p(n_{\sigma(1)},n_{\sigma(2)},
\ldots,n_{\sigma(k)}).
\end{equation}
Let $\Pi$ be
the partition of $[n]$ with $\bfn(\Pi) = (n_1,\ldots,n_k)$ such that
\[
\Pi\bigl([n]\bigr) = (1,2,\ldots,k,1,\ldots,1,2,\ldots,2,\ldots,k,\ldots,k),
\]
where after the first $k$ elements $1,2,\ldots,k$, $i$ is repeated
$n_i-1$ times for all $i=1,\ldots,k$. Then,
\[
p(\bfn) = \prod_{i=2}^k p_i(
\bfone_{(i-1)})\times\prod_{i=k}^{n-1}
p_{\Pi(i+1)}\bigl(\bfn(\Pi_i)\bigr),
\]
where $\bfone_{(j)}$ is the vector of
length $j$ whose elements are all $1$'s.

%
%f1 #&#
\begin{figure*}[b]

\includegraphics{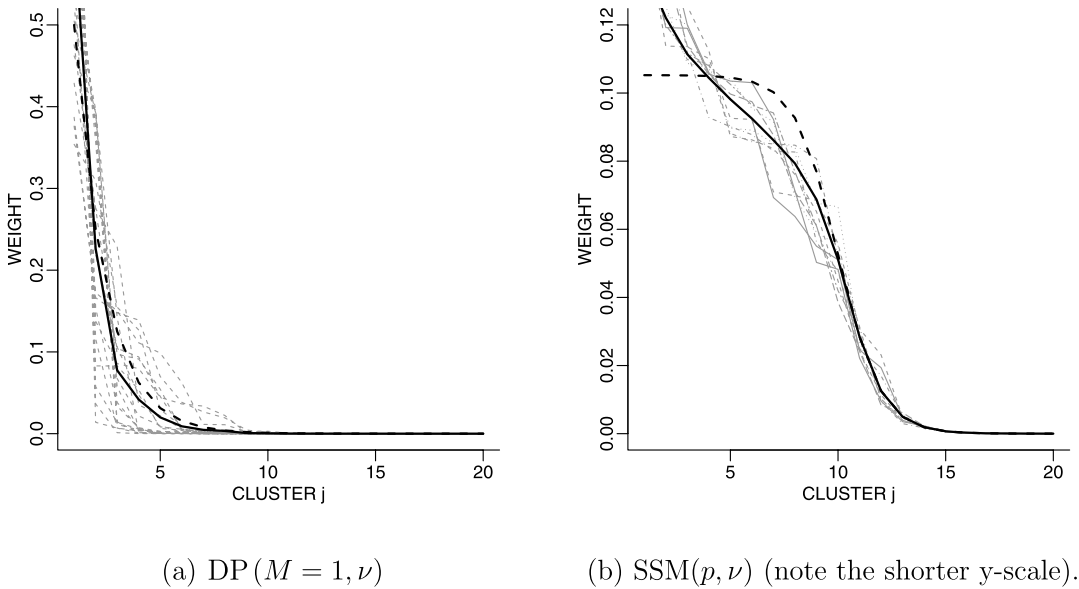}

%(a) $DP(M=1,\nu)$
%(b) $\SSM(p,\nu)$ (note the shorter y-scale).
\caption{The lines in each panel show 10 draws
$\bP\sim p(\bP)$ for the DP (left) and for the SSM defined
in (\protect\ref{eqex}) below (right).
The $P_h$ are defined for integers $h$ only.
We connect them to a line for presentation only.
Also, for better presentation we plot the \textup{sorted} weights.
The thick line shows the prior mean. For comparison, a dashed
thick line plots the prior mean of the \textup{unsorted} weights.
Under the DP the sorted and unsorted prior means are almost
indistinguishable.}
\label{figw}
\end{figure*}

Now consider a partition $\Omega$ of $[n]$ with $\bfn(\Omega) =
(n_{\sigma(1)},n_{\sigma(2)},\ldots,n_{\sigma(k)})$ such that
\begin{eqnarray*}
\Omega\bigl([n]\bigr) &=& \bigl(1,2,\ldots,k,\sigma^{-1}(1),\ldots,\sigma^{-1}(1),\\
&&\hspace*{6pt}
\sigma^{-1}(2),\ldots,\sigma^{-1}(2),
\ldots,\\
&&\hspace*{60pt}\sigma^{-1}(k),\ldots,\sigma^{-1}(k)\bigr),
\end{eqnarray*}
where after the first $k$ elements $1,2,\ldots,k$, $\sigma^{-1}(i)$ is
repeated $n_i-1$ times for all $i=1,\ldots,k$. Then,
\begin{eqnarray*}
&&
p(n_{\sigma(1)},n_{\sigma(2)},\ldots,n_{\sigma(k)}) \\
&&\quad = \prod
_{i=2}^k p_i(
\bfone_{(i-1)})\times\prod_{i=k}^{n-1}
p_{\Omega(i+1)}\bigl(\bfn(\Omega_i)\bigr)
\\[-2pt]
&&\quad = \prod_{i=2}^k p_i(
\bfone_{(i-1)})\times\prod_{i=k}^{n-1}
p_{\sigma^{-1}(\Omega(i+1))} \bigl(\sigma\bigl(\bfn(\Omega_i)\bigr)\bigr)
\\[-2pt]
&&\quad = \prod_{i=2}^k p_i(
\bfone_{(i-1)})\times\prod_{i=k}^{n-1}
p_{\Pi(i+1)}\bigl(\bfn(\Pi_i)\bigr)
\\[-2pt]
&&\quad = p(n_1,\ldots,n_k),
\end{eqnarray*}
where the second equality follows from (\ref{eqsuffeppf}). This
completes the proof of the sufficient direction.

Finally, we show that every EPPF $p$ satisfies (\ref{eqeppfdef})
and (\ref{eqsuffeppf}). By Lemma~\ref{lemm1}, every EPPF satisfies
(\ref{eqeppfdef}). Condition (\ref{eq9a}) is true by the definition of
an EPPF, which includes the condition of symmetry in its arguments. And
(\ref{eq9a}) implies (\ref{eqsuffeppf}).
\end{pf}

\citet{FortiniLadelli00} prove results related to Theorem
\ref{theo1}. They provide sufficient conditions for a system of
predictive distributions $p(X_{n}\mid X_1,\ldots,X_{n-1})$,
$n=1,\ldots\,$, of a sequence of random variables $(X_i)$ that imply
exchangeability. The relation between these conditions and Theorem~\ref{theo1}
becomes apparent by constructing a sequence $(X_i)$ that
induces a $p$-distributed random partition of $\mathbb{N}$. Here, it is
implicitly assumed the mapping of $(X_i)$ to the only partition such
that $i,j\in\mathbb{N}$ belongs to the same subset if and only if
$X_i=X_j$.

A second more general example, which extends the predictive structure
considered in Corollary~\ref{coro1}, includes the\vadjust{\goodbreak} so-called Gibbs random
partitions.\break Within this class of models
%
%e13 #&#
\begin{equation}
\label{Gibbs} p(n_1, n_2, \ldots, n_k) =
V_{n,k} \prod_{i=1}^k
W_{n_i},
\end{equation}
where $(V_{n,k})$ and $(W_{n_i})$ are sequences of positive real
numbers. In this case the predictive probability of a novel species
is a function of the sample size $n$ and of the number of observed
species $k$. See \citet{Lijoi2007} for related distributional results
on Gibbs type models. \citet{GnedinPitman06} obtained sufficient
conditions for the sequences $(V_{n,k})$ and $(W_{n_i})$, which imply
that $p$ is an EPPF.

%s3 #&#
\section{SSMs Beyond the DP} \label{seclorenzo}

%s3.1 #&#
\subsection{\texorpdfstring{The $\SSM(p,\nu)$}{The SSM(p, nu)}}\label{secdirect}

We know that an SSM with a nonlinear PPF, that is, $p_j$ different from
the PPF of a DP, cannot be described as a function $p_j \propto
f(n_j)$ of $n_j$ only. It must be a more complicated function
$f(\bfn)$. Alternatively, one could try to define an EPPF and deduce
the implied PPF. But directly specifying a symmetric function $p(\bfn)$
such that it complies with (\ref{eqeppf}) is difficult. As a third
alternative we propose to consider the weights $\bP= \{P_{h},
h=1,2,\ldots\}$ in (\ref{eqssp}).
Figure~\ref{figw}(a) illustrates $p(\bP)$ for a DP model. The sharp
decline is typical.\vadjust{\goodbreak} A~few large weights account for most of the
probability mass. The stick breaking construction for a DP prior with
total mass $\th$ implies $E(P_h) = \th^{h-1} (1+\th)^{-h}$. Such
geometrically decreasing mean weights are inappropriate to describe
prior information in many applications. The weights can be interpreted
as asymptotic relative cluster sizes. A~typical application of the DP
prior is, for example, a partition of patients in a clinical study into
clusters. However, if clusters correspond to disease subtypes defined
by variations of some biological process, then one would rather expect
a number of clusters with a priori comparable size. Many small clusters
with very few patients are implausible and would also be of little
clinical use. This leads us to propose the use of alternative SSMs.

Figure~\ref{figw}(b) shows an alternative probability mo\-del~$p(\bP)$.
There are many ways to define $p(\bP)$; we consider, for
$h=1,2,\ldots\,$,
\[
P_h \propto u_h \quad\mbox{or}\quad P_h =
\frac{u_h}{\sum_{i=1}^\infty u_i},
\]
where $u_h$ are independent and nonnegative random variables with
%
%e14 #&#
\begin{equation}
\label{condition1} \sum_{i=1}^\infty
u_i < \infty\quad \mbox{a.s.}
\end{equation}
A sufficient condition for (\ref{condition1}) is
%
%e15 #&#
\begin{equation}
\label{condition2} \sum_{i=1}^\infty
E(u_i) < \infty
\end{equation}
by the monotone convergence theorem. Note that when the unnormalized
random variables $u_h$ are defined as the sorted atoms of a
nonhomogeneous Poisson process on the positive real line, under mild
assumptions, the above $(P_h)$ construction coincides with the
Poisson--Kingman models. \citet{FergusonKlass72} provide a detailed
discussion on the outlined mapping of a Poisson process into a
sequence of unnormalized positive weights. In this particular case the
mean of the Poisson process has to satisfy minimal requirements
(see, e.g., \citecs{Pitman03}) to ensure that the sequence
$(P_i)$ is well defined.

As an illustrative example in the following discussion, we define, for
$h=1,2,\ldots\,$,
%
%e16 #&#
\begin{eqnarray}\label{eqex}
P_h \propto e^{X_h} \hspace*{140pt}\nonumber\\[-8pt]\\[-8pt]
&&\eqntext{\mbox{with } X_h \sim N
\bigl(\log\bigl(1-\bigl\{1+e^{b- a h}\bigr\}^{-1}\bigr),
\sigma^2 \bigr),}
\end{eqnarray}
where $a, b, \sigma^2$ are positive constants. The existence of such
random probabilities is guaranteed by (\ref{condition2}), which is easy
to check.

The S-shaped nature of the random distribution (\ref{eqex}), when
plotted against $h$, distinguishes it from the DP model. The first few
weights are a priori of equal size (before sorting). This is in
contrast to the stochastic ordering of the DP and the Pitman--Yor
process in general. In panel (a) of Figure~\ref{figw} the prior mean
of the sorted and unsorted weights is almost indistinguishable,
because the prior already implies strong stochastic ordering of the
weights.

The prior in Figure~\ref{figw}(b) reflects prior information of an
investigator who believes that there should be around 5 to 10 clusters
of comparable size in the population. This is in sharp contrast to the
(often implausible) assumption of one large dominant cluster and
geometrically smaller clusters that is reflected in panel (a). Prior
elicitation can exploit such readily interpretable implications of the
prior choice to propose models like~(\ref{eqex}).

We use $\SSM(p,\nu)$ to denote a SSM defined by $p(\bP)$ for the
weights $P_h$ and $m_h \iid\nu$. The attraction of
defining the SSM through $\bP$ is that by (\ref{eqssp}) any joint
probability model $p(\bP)$ such that\break $P(\sum_h P_h=1)$ defines a proper
SSM. There are no additional constraints as for the PPF $p_j(\bfn)$ or
the EPPF $p(\bfn)$. However, we still need the implied PPF to implement
posterior inference and also to understand the implications of the
defined process. Thus, a practical use of this second approach requires
an algorithm to derive the PPF starting from an arbitrarily defined
$p(\bP)$.

%s3.2 #&#
\subsection{An Algorithm to Determine the PPF}
\label{secppfalgo}

Recall definition (\ref{eqssp}) for an SSM random probability measure.
Assuming a proper SSM, we have
%
%e17 #&#
\begin{equation}\label{eqSSM}
G = \sum_{h=1}^{\infty} P_h
\delta_{m_h}.
\end{equation}
Let $\Pb=(P_h, h \in\bbN)$ denote the sequence of weights. Recall the
notation $\Xt_j$ for the $j$th unique value in the SSS $\{X_i, i =1,
\ldots, n\}$. The algorithm requires indicators that match the $\Xt_j$
with the $m_h$, that is, that match the clusters in the partition with the
point masses of the SSM. Let $\pi_j=h$ if $\Xt_j=m_h$,
$j=1,\ldots,k_n$. In the following discussion it is important that the
latent indicators $\pi_j$ are only introduced up to $j=k$. Conditional
on $m_h$, $h \in\bbN$ and $\Xt_j$, $j\in\bbN$, the indicators $\pi_j$
are deterministic. After marginalizing with respect to the $m_h$ or
with respect to the $\Xt_j$, the indicators become latent variables.
Also, we use cluster membership indicators $s_i=j$ for $X_i=\Xt_j$ to
simplify notation.\vadjust{\goodbreak} We use the convention of labeling clusters in the
order of appearance, that is, $s_1=1$ and $s_{i+1} \in
\{1,\ldots,k_i,k_i+1\}$.

In words, the algorithm proceeds as follows. We write the desired PPF
$p_j(\bfn)$ as an expectation of the conditional probabilities
$p(X_{n+1}=\Xt_j \mid\bfn, \pi, \Pb)$. The expectation is with respect
to $p(\Pb,\pi\mid\bfn)$. Next we approximate the integral with
respect to $p(\Pb,\pi\mid\bfn)$ by a weighted Monte Carlo average
over samples $(\Pbell,\piell) \sim p(\Pbell) p(\piell\mid\Pbell)$
from the prior. Note $\pi$ and $\Pb$ together define the size-biased
permutation of $(P_j)$,
\[
\tilde{P}_j = P_{\pi_j},\quad j=1,2,\ldots.
\]
The size-biased permutation $(\tilde{P}_j)$ of $(P_j)$ is a resampled
version of $(P_j)$ where sampling is done with probability proportional
to $P_j$ and without replacement. Once the sequence $(P_j)$ is
simulated, it is computationally straightforward to get
$(\tilde{P}_j)$. Note also that the properties of the random partition
can be characterized by the distribution on $\Pb$ only. The point
masses $m_h$ are not required.

Using the cluster membership indicators $s_i$ and the size-biased
probabilities $\tilde{P}_j$, we write the desired PPF as
%
%e18 #&#
\begin{eqnarray}\label{eqppf}\qquad
p_j(\bfn) &=& p(s_{n+1}=j \mid\bfn)
\nonumber
\\
& = & \int p(s_{n+1}=j \mid\bfn, \tilde{\Pb}) p( \tilde{\Pb} \mid\bfn)
\dd\tilde{\Pb}
\nonumber\\[-8pt]\\[-8pt]
&\propto& \int p(s_{n+1}=j \mid\bfn, \tilde{\Pb}) p(\bfn\mid\tilde{
\Pb}) p(\tilde{\Pb}) \dd\tilde{\Pb}
\nonumber
\\
& \approx& \frac1L \sum p\bigl(s_{n+1}=j \mid\bfn,\tPbell\bigr) p\bigl(\bfn
\mid\tPbell\bigr).\nonumber
\end{eqnarray}

The Monte Carlo sample $\tPbell$ or, equivalently,\break $(\Pbell,\pibell)$,
is obtained by first generating $\Pbell\sim p(\Pb)$ and then
$p(\piell_j=h \mid\Pbell, \piell_1,\ldots,\piell_{j-1}) \propto
\Pell_h$, $h \notin\{\piell_1,\ldots,\piell_{j-1}\}$. In actual
implementation the elements of $\Pbell$ and $\pibell$ are only
generated as and when needed.

The terms in the last line of (\ref{eqppf}) are easily evaluated. The
first factor is given as predictive cluster membership probabilities
%
%e19 #&#
\begin{eqnarray}
\label{eqppfi}
&&
p(s_{n+1}=j \mid\bfn,\tPb)\nonumber\\[-8pt]\\[-8pt]
&&\quad = \cases{
\tilde{P}_j, & $j=1,\ldots,k_n$,
\vspace*{2pt}\cr
\displaystyle \Biggl(1-\sum
_{j=1}^{k_n} \tilde{P}_j\Biggr), &
$j=k_n+1$.}\nonumber
\end{eqnarray}
The second factor is evaluated as
\[
p(\bfn\mid\tPb) = \prod_{j=1}^k
\tilde{P}_j^{n_j-1} \prod_{j=1}^{k-1}
\Biggl(1-\sum_{i=1}^{j}
\tilde{P}_i\Biggr).
\]
Note that the second factor coincides with the previously mentioned
[cf. expression (\ref{eqpartialcar})] Pitman's representation result
for partially exchangeable partitions.

%
%f2 #&#
\begin{figure*}

\includegraphics{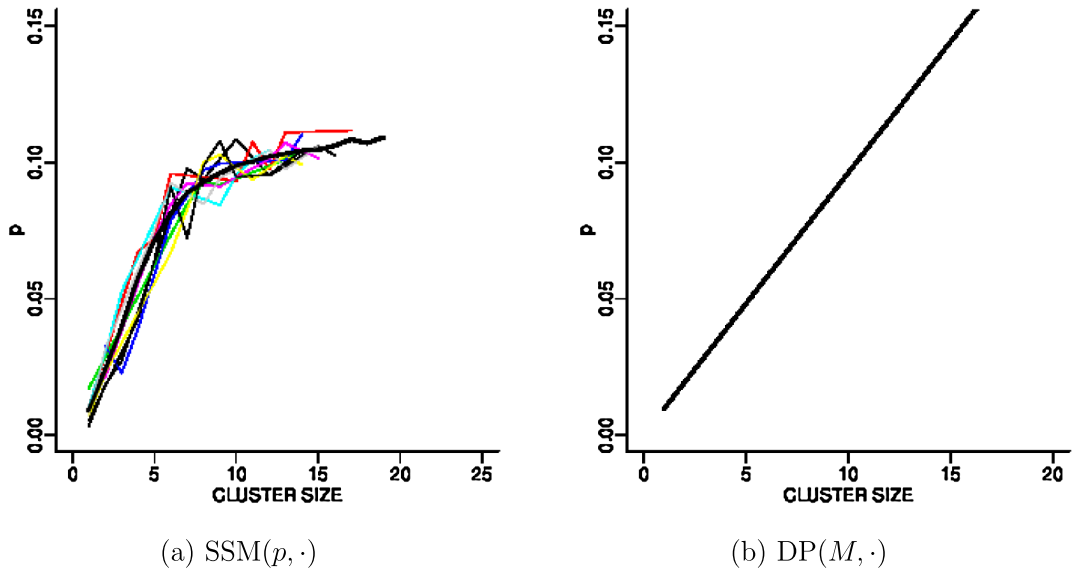}

%(a) $\SSM(p,\cdot)$ & (b) $\DP(M,\cdot)$
\caption{Panel \textup{(a)} shows the PPF (\protect\ref{eqppfi}) for a
random probability measure $G \sim\SSM(p,\nu)$, with $P_h$ as in
(\protect\ref{eqex}). The thick line plots $p(s_{n+1}=j \mid\bs)$
against $n_j$, averaging over multiple simulations. In each simulation
we used the same simulation truth to generate $\bs$ and stop simulation
at $n=100$. The 10 thin lines show $p_j(\bfn)$ for 10 simulations with
different $\bfn$. In contrast, under the DP Polya urn the curve is a
straight line and there is no variation across simulations [panel
\textup{(b)}].}\vspace*{-3pt} \label{figppf}
\end{figure*}

%f3 #&#
\begin{figure*}[b]
\vspace*{-3pt}

\includegraphics{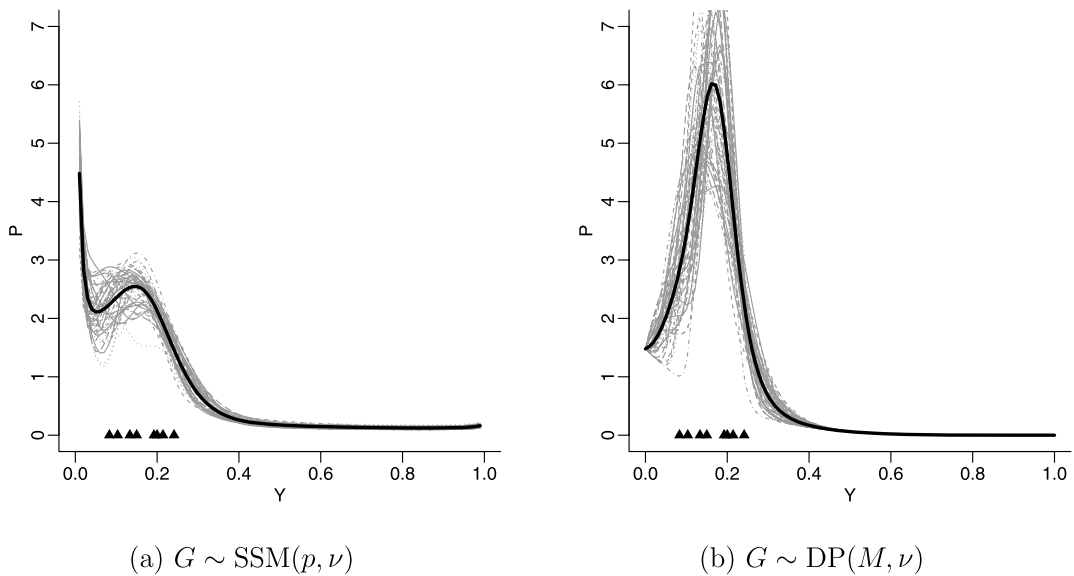}

%(a) $G \sim\SSM(p,\nu)$ &
%(b) $G \sim\DP(M, \nu)$
\caption{Posterior estimated sampling model \mbox{$\Fbar= E(F \mid\mathrm{data}) = p(y_{n+1}
\mid\mathrm{data})$} under the $\SSM(p,\nu)$ prior and a comparable DP
prior. The triangles along the x-axis show the data.}
\label{figfbar}
\end{figure*}

Figure~\ref{figppf} shows an example. The figure plots\break $p(s_{i+1}=j
\mid\bs)$ against cluster size $n_j$. In contrast, the DP Polya urn
would imply a straight line. The plotted probabilities are averaged
with respect to all other features of $\bs$, in particular, the
multiplicity of cluster sizes, etc. The figure also shows probabilities
(\ref{eqppfi}) for specific simulations.

%s3.3 #&#
\subsection{A Simulation Example}
\label{secex}

Many data analysis applications of the DP prior are based on DP
mixtures of normals as models for a random probability measure $F$.
Applications include density estimation, random effects distributions,
generalizations of a probit link, etc. We consider a stylized example
that is chosen to mimic typical features of such models.

%f4 #&#
\begin{figure*}

\includegraphics{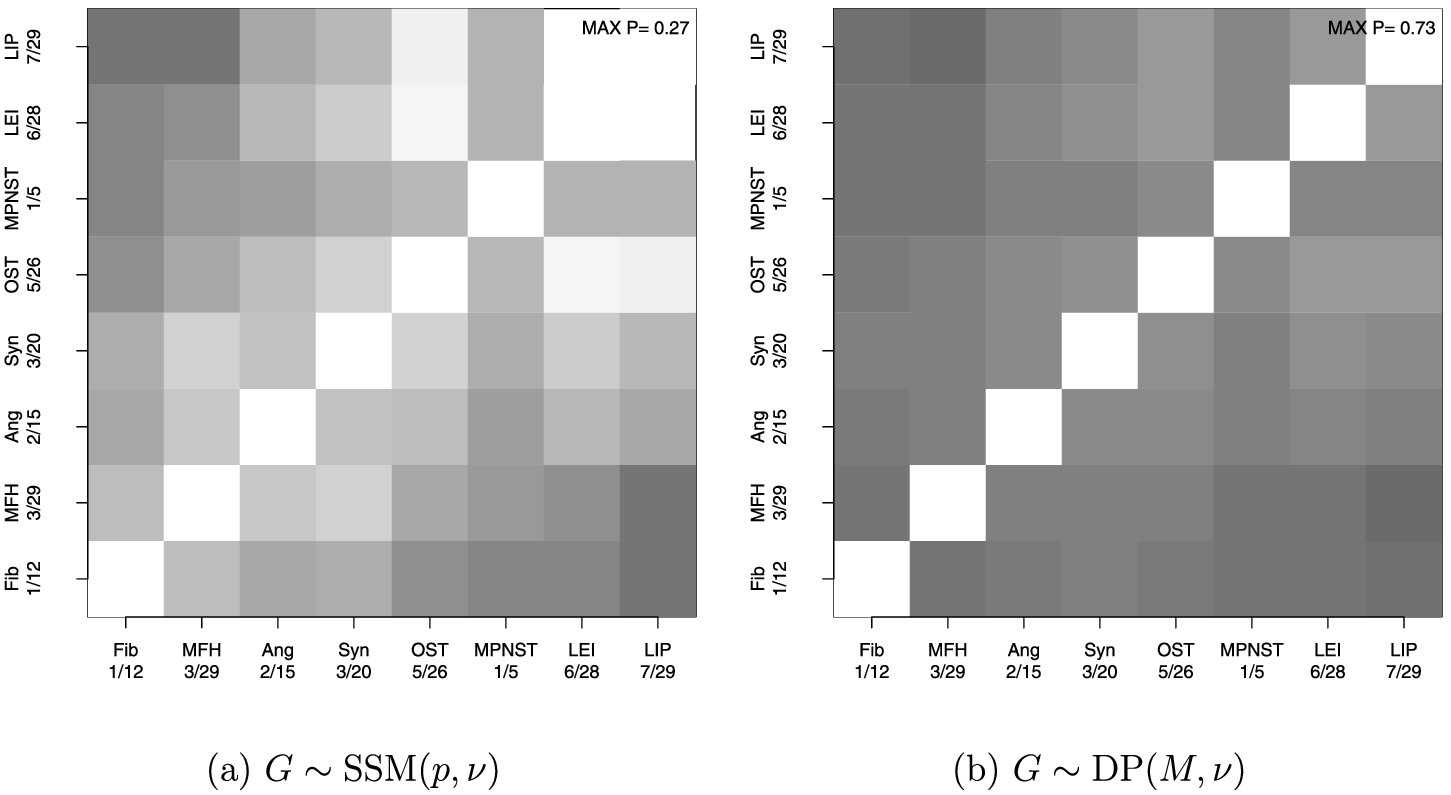}

%(a) $G \sim\SSM(p,\nu)$ &
%(b) $G \sim\DP(M, \nu)$
\caption{Co-clustering probabilities
$p(s_i=s_j \mid\mathrm{data})$ under the two prior models.}
\label{figsij}
\end{figure*}

In this section we show posterior inference conditional on the data
set $(y_1, y_2,\ldots,y_9)= (-4,-3,-2,\break\ldots,4)$. The use of these data
highlights the differences in posterior inference between the SSM and
DP priors. Assume $y_i \iid F$, with a semi-parametric mixture of
normal prior on $F$,
\[
y_i \iid F\quad \mbox{with } F(y_i) = \int N
\bigl(y_i; \mu,\sigma^2\bigr) \dd G\bigl(\mu,
\sigma^2\bigr).
\]
Here $N(x; m,s^2)$ denotes a normal distribution with moments
$(m,s^2)$ for the random variable $x$. We estimate $F$ under two
alternative priors,
\[
G \sim\SSM(p,\nu) \quad\mbox{or}\quad G \sim\DP(M,\nu).
\]
The distribution $p$ of the weights for the $\SSM(p,\cdot)$ prior is
defined as in (\ref{eqex}). The total mass parameter $M$ in the DP
prior is fixed to match the prior mean number of clusters, $E(k_n)$,
implied by (\ref{eqex}). We find $M=2.83$. Let $\Ga(x; a,b)$ indicate
that the random variable $x$ has a Gamma distribution with shape
parameter $a$ and inverse scale parameter $b$. For both prior models we
use
\[
\nu\bigl(\mu,1/\sigma^2\bigr)= N\bigl(x; \mu_0, c
\sigma^2\bigr) \Ga\bigl(1/\sigma^2; a/2,b/2\bigr).
\]
We fix $\mu_0=0$, $c=10$ and $a=b=4$. The model can alternatively be
written as $y_i \sim N(\mu_i,\sigma_i^2)$ and
$X_i=(\mu_i,1/\sigma^2_i) \sim G$.

Figures~\ref{figfbar} and~\ref{figsij} show some inference summaries.
Inference is based on Markov chain Monte Carlo (MCMC) posterior
simulation with $1000$ iterations. Posterior simulation is for
$(s_1,\ldots,s_n)$ only. The clus\-ter-specific parameters $(\tilde
\mu_j,\tilde\sigma_j^2)$, $j=1,\ldots,k_n$, are analytically
marginalized. One of the transition probabilities (Gibbs sampler) in
the MCMC requires the PPF under $\SSM(p,\nu)$. It is evaluated using
(\ref{eqppf}).\vadjust{\goodbreak}

Figure~\ref{figfbar} shows the posterior estimated sampling
distributions $F$. The figure highlights a limitation of the DP prior.
The single total mass parameter $M$ controls both, the number of
clusters and the prior precision. A small value for $M$ favors a small
number of clusters and implies low prior uncertainty. Large $M$ implies
the opposite. Also, we already illustrated in Figure~\ref{figw} that
the DP prior implies stochastically ordered cluster sizes, whereas the
chosen SSM prior allows for many approximately equal size clusters. The
equally spaced grid data $(y_1,\ldots,y_n)$ implies a likelihood that
favors a moderate number of approximately equal size clusters. The
posterior distribution on the random partition is shown in Figure
\ref{figsij}. Under the SSM prior the posterior supports a moderate
number of similar size clusters. In contrast, the DP prior shrinks the
posterior toward a few dominant clusters. Let $n_{(1)} \equiv
\max_{j=1,\ldots,k_n} n_j$ denote the leading cluster size. Related
evidence can be seen in the marginal posterior distribution (not shown)
of $k_n$ and $n_{(1)}$. We find $E(k_n \mid\mathrm{data})=6.4$ under the
SSM model versus $E(k_n \mid\mathrm{data})=5.1$ under the DP prior. The
marginal posterior modes are $k_n=6$ under the SSM prior and $k_n=5$
under the DP prior. The marginal posterior modes for $n_{(1)}$ is
$n_{(1)}=2$ under the SSM prior and $n_{(1)}=3$ under the DP prior.

%s3.4 #&#
\subsection{Analysis of Sarcoma Data}

We analyze data from of a small phase II clinical trial for sarcoma
patients that was carried out in the M.~D. Anderson Cancer Center. The
study was designed to assess efficacy of a treatment for sarcoma
patients across different subtypes. We consider the data accrued for
$8$ disease subtypes that were classified as having overall
intermediate prognosis, as presented in Table~\ref{tabdata}. Each
table entry indicates the total number of patients for each sarcoma
subtype and the number of patients who reported a treatment success.
See further discussion in \citet{leonnoveloetal12}.

%t1 #&#
\begin{table}
\tabcolsep=0pt
\caption{Sarcoma data. For each disease subtype (top row)\break we report
the total
number of patients and the number\break of treatment successes. See
Le{\'o}n-Novelo et al. (\citeyear{leonnoveloetal12})\break
for a discussion of disease subtypes}\label{tabdata}
\begin{tabular*}{\tablewidth}{@{\extracolsep{\fill}}lcccccccc@{}}
\hline
\textbf{Sarcoma} &
\textbf{LEI} & % Leiomyosarcoma
\textbf{LIP} & % Liposarcoma
\textbf{MFH} &
\textbf{OST} & % steosarcoma &
\textbf{Syn} & % ovial &
\textbf{Ang} & % iosarcoma &
\textbf{MPNST} &
\textbf{Fib} \\ %rosarcoma \\
\hline
& $6/28$ & $7/29$ & $3/29$ & $5/26$ & $3/20$ & $2/15$ & $1/5$ & $1/12$\\
\hline
\end{tabular*}
\end{table}
%

%f5 #&#
\begin{figure*}

\includegraphics{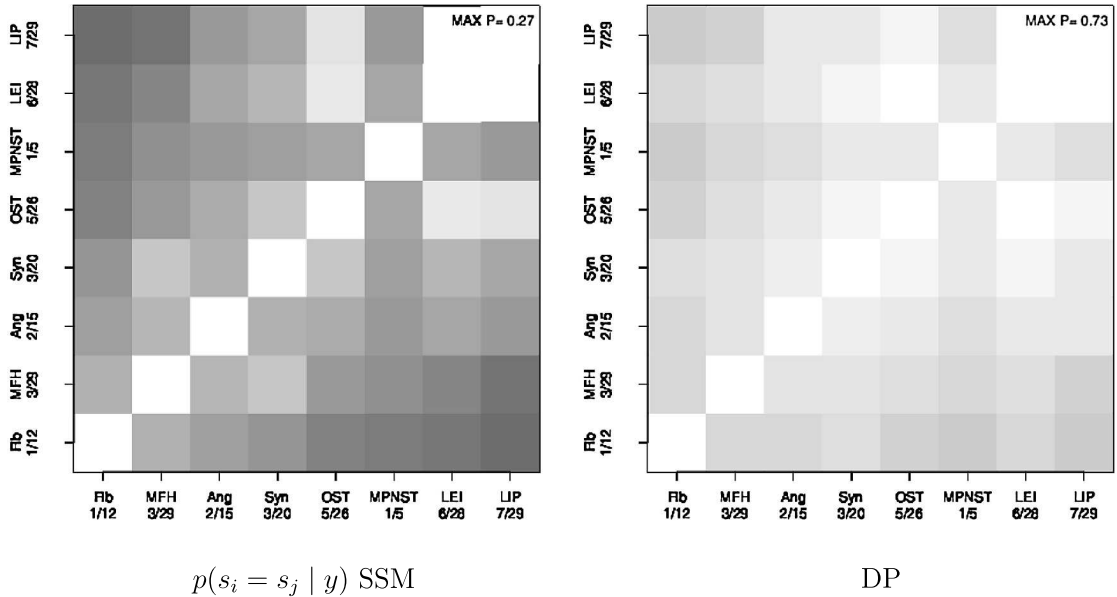}

%$p(s_i=s_j \mid y)$ SSM & DP
\caption{Posterior probabilities of pairwise co-clustering, $p_{ij} =
p(s_i =
s_j \mid y)$. The grey scales in the two panels are scaled as black for
$p_{ij}=0$ to white for $p_{ij}=\max_{r,s} p_{rs}$. The maxima are
indicated in the right top of the plots.}\label{figpairwise}
\end{figure*}

One limitation of these data is the small sample size, which prevents
separate analysis for each disease subtype. On the other hand, it is
not clear that we should simply treat the subtypes as exchangeable. We
deal with these issues by modeling each table entry as a binomial
response and adopt a hierarchical framework for the success
probabilities. The hierarchical model includes a random partition of
the subtypes. Conditional on a given partition, data across all
subtypes in the same cluster are pooled, thus allowing more precise
inference on the common success probabilities for all subtypes in this
cluster. We consider two alternative models for the random partition,
based on a $\DP(M,\nu)$ prior versus a $\SSM(p,\nu)$ prior.
Specifically, we consider the following models:
\begin{eqnarray*}
y_i|\pi_i & \sim& \operatorname{Bin}(n_i,
\pi_i),
\\
\pi_i|G & \sim& G,
\\
G & \sim& \DP(M, \nu) \quad\mbox{or}\quad \SSM(p, \nu),
\end{eqnarray*}
where $\nu$ is a diffuse probability measure on $[0,1]$ and $p$ is
again defined as in (\ref{eqex}).

The hierarchical structure of the data and the aim of clustering
subpopulations in order to achieve borrowing of strength is in
continuity with a number of applied contributions. Several of these,
for instance, are meta analyses of medical studies
(Berry and\break Christensen, \citeyear{berrychristensen79}), with subpopulations defined by medical
institutions or by clinical trials. In most cases the application of
the DP is chosen for computational advantages and (in some cases) due
to the easy implementation of strategies for prior specification
\citep{LIU96}. With a small number of studies, as in our example, ad
hoc construction of alternative SSM combines hierarchical modeling
with advantageous posterior clustering. The main advantage is the
possibility of avoiding the exponential decrease typical of the ordered
DP atoms.

In this particular analysis, we used $M=2.83$ and chose $\nu$ to be
the $\operatorname{Beta}(0.15,0.85)$ distribution, which was designed to match the
prior mean of the observed data and has prior equivalent sample size
of~$1$. The total mass $M=2.83$ for the DP prior was selected to
achieve matching prior expected number of clusters under the two
models. The DP prior on $G$ favors the formation of large clusters
(with matched prior mean number of clusters) which leads to less
posterior shrinkage of cluster-specific means. In contrast, under the
SSM prior the posterior puts more weight on several smaller clusters.

Figure~\ref{figpairwise} shows the estimated posterior probabilities
of pairwise co-clustering for model (\ref{eqex}) in the left panel
and for the DP case (right panel). Clearly, compared to the DP
model, the chosen SSM induces a posterior distribution with more
clusters, as reflected in the lower posterior probabilities
$p(s_i=\break s_j\mid y)$ for all $i,j$.

%f6 #&#
\begin{figure*}

\includegraphics{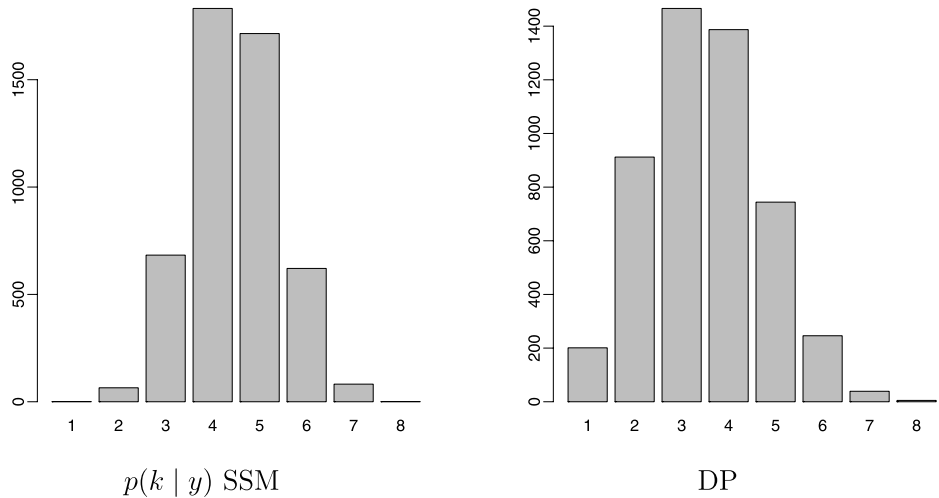}

%$p(k \mid y)$ SSM & DP
\caption{Posterior distribution on the number of clusters.} \label{figncl}
\end{figure*}

%f7 #&#
\begin{figure*}[b]

\includegraphics{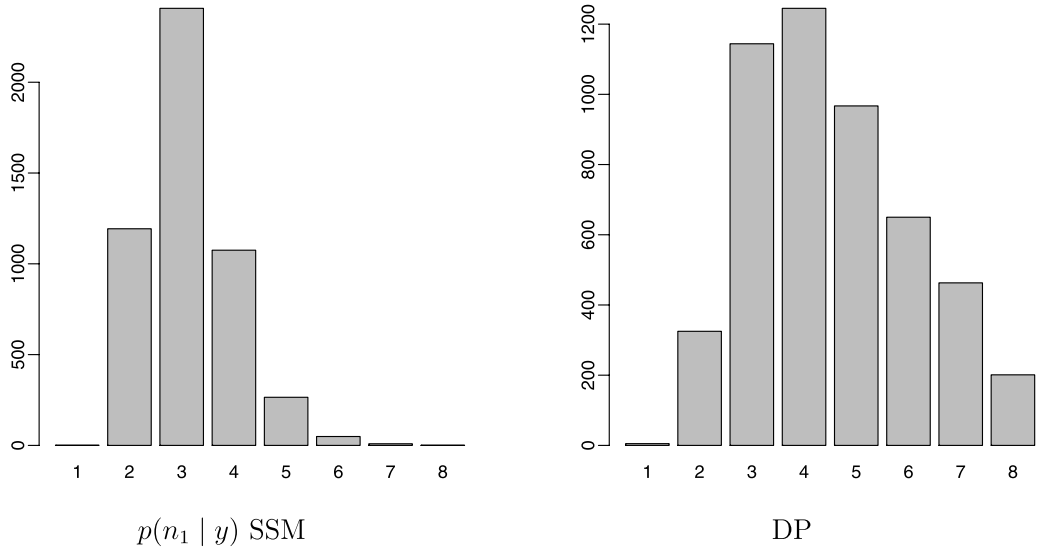}

%$p(n_1 \mid y)$ SSM & DP
\caption{Posterior distribution on the size of the largest cluster.}
\label{figlargest}
\end{figure*}

Figure~\ref{figncl} shows the posterior distribution of the number of
clusters under the SSM and DP mixture models. Under the DP (right
panel) includes high probability for a single cluster, $k=1$, with
$n_1=8$. The high posterior probability for few large clusters also
implies high posterior probabilities $\phat_{ij}$ of co-clustering.
Under the SSM (left panel) the posterior distribution on $\rho$ retains
substantial uncertainty. Finally, the same pattern is confirmed in the
posterior distribution of sizes of the largest cluster, $p(n_1 \mid
y)$, shown in Figure~\ref{figlargest}. The high posterior probability
for a single large cluster of all $n=8$ sarcoma subtypes seems
unreasonable for the given data.

%s4 #&#
\section{Discussion}

We have reviewed alternative definitions of SSMs. We also reviewed the
fact that all SSMs with a PPF of the form $p_j(\bfn) = f(n_j)$ must
necessarily be a linear function of $n_j$ and provided a new elementary
proof. In other words, the PPF $p_j(\bfn)$ depends on the current data
only through the cluster sizes. The number of clusters and any other
aspect of the partition $\Pi_n$ do not change the prediction. This is
an excessively simplifying assumption for most data analysis problems.

We provide an alternative class of models that allows for more general
PPFs. These models are obtained by directly specifying the
distribution of unnormalized weights $u_h$. The proposed approach for
defining SSMs allows the incorporation of the desired qualitative
properties concerning the decrease of the ordered clusters
cardinalities. This flexibility comes at the cost of additional
computation required to implement the algorithm described in
Section~\ref{secppfalgo}, compared to the standard approaches under
DP-based models. Nevertheless, the benefits obtained in the case of
data sets that require more flexible models compensate the increase in
computational effort. A different strategy for constructing discrete
random distributions has been discussed in \citet
{trippafavaro12}. In
several applications, the scope for which SSMs are to be used suggests
these \textit{desired qualitative properties}. Nonetheless, we see the
definition of a theoretical framework supporting the selection of a SSM
as an open problem.

R code for an implementation of posterior inference under the proposed
new model is available at \url{http://math.utexas.edu/users/pmueller/}.

\section*{Acknowledgments}

%Jaeyong Lee was supported by the Basic Science Research Program through the
%National Research Foundation of Korea (NRF) funded by the Ministry of
%Education, Science and Technology (20090075171).
Jaeyong Lee was supported by the National Research Foundation of Korea
(NRF) grant funded by the Korea government (MEST) (No. 2011-0030811).
Fernando Quintana was
supported by Grant\break FONDECYT 1100010. Peter M\"uller was partially
funded by Grant NIH/NCI CA075981.

%suskaldyti doi

% imsref loaded by lrinkeviciute, 2012-10-02 15:51:34
% imsref loaded by lrinkeviciute, 2012-10-03 08:20:47
%

\end{document}